\documentclass[twocolumn,preprintnumbers,amsmath,prc,superscriptaddress,amssymb]{revtex4}

\usepackage{graphicx}
\usepackage{dcolumn}
\usepackage{bm}
\usepackage{amsmath}
\usepackage{lipsum}
\usepackage{slashed}
\usepackage{float}
\usepackage{xcolor}

\begin{document}

\title{A second look to the Polyakov Loop Nambu-Jona-Lasinio model\\ at finite baryonic density}

\date{\today}


\author{O. Ivanytskyi}
\email{oivanytskyi@usal.es}
\affiliation{Department of Fundamental Physics, University of Salamanca, Plaza de la Merced S/N E-37008, Salamanca, Spain}

\author{M. \'Angeles P\'erez-Garc\'ia}
\email{mperezga@usal.es}
\affiliation{Department of Fundamental Physics, University of Salamanca, Plaza de la Merced S/N E-37008, Salamanca, Spain}

\author{V. Sagun}
\email{violetta.sagun@uc.pt}
\affiliation{CFisUC, Department of Physics, University of Coimbra, Rua Larga P-3004-516, Coimbra, Portugal}

\author{C. Albertus}
\email{albertus@usal.es}
\affiliation{Department of Fundamental Physics, University of Salamanca, Plaza de la Merced S/N E-37008, Salamanca, Spain}

\date{\today}

\begin{abstract}

We revisit the Polyakov Loop coupled Nambu-Jona-Lasinio model that maintains the Polyakov loop dynamics in the limit of zero temperature. This is of interest  for astrophysical applications in the interior of neutron stars. For this purpose we re-examine the form of the potential for the deconfinement order parameter at finite baryonic densities. Since the modification of this potential at any temperature is formally equivalent to assigning a baryonic charge to gluons,  we develop a more general formulation of the present model that cures this spurious effect and is normalized to match the asymptotic behaviour of the QCD equation of state given by $\mathcal{O}(\alpha_s^2)$  and partial $\mathcal{O}(\alpha_s^3\ln^2\alpha_s)$ perturbative results.


\end{abstract}

\keywords{}

\maketitle

\section{Introduction}

During the last decades the properties of matter at extreme conditions have been intensively studied both theoreti\-cally and experimentally. Phase transformations governed by strong interaction, i.e. the deconfinement of colour degrees of freedom and restoration of chiral symmetry, belong to the most important subjects of these studies. The interest to the mentioned phenomena is stimulated by experiments on collisions of ultra relativistic heavy ions performed at facilities RHIC and LHC, which have already yielded signals of the quark-gluon plasma (QGP) existence \cite{QGP}. Moreover, future experimental programs planned in FAIR GSI, NICA JINR and J-PARK are in a dire need of information about the phase diagram of Quantum Chromodynamics (QCD), being a modern theory of strong interaction. 

Another practical need of information about phase transformations in strongly interacting matter is related to the possible existence of hybrid compact stars with a quark core \cite{HS1,HS2}. Furthermore, the recently predicted sudden increase of frequency of gravitational waves emitted in mergers of such hybrid stars opens a remarkable possibility of their detection \cite{HS3} but, on the other hand, it still requires further clarification of some details of the quark-hadron transition.

A complete knowledge about this transition can be reached only within QCD, which, even despite the tremendous efforts documented in the literature, is not satisfactorily solved due to its nonperturbative character. At the same time, a significant progress toward the understanding the phase structure of QCD has been achieved during the last years. First principle calculations on discrete space-time lattices provided access to the equation of state (EoS) in the regime of high temperatures $T$ and limited baryonic chemical potentials $\mu_B$ \cite{lQCD1,lQCD2,lQCD3,lQCD4}. Taylor series expansion \cite{lQCD5,lQCD6,lQCD7}, re-weighting techniques \cite{lQCD8,lQCD9,lQCD10} and analytical continuation from imaginary chemical potentials \cite{lQCD11,lQCD12,lQCD13,lQCD14,lQCD15} extended the applicability range of lattice QCD up to $\frac{\mu_B}{T}\le 3$. Further extension remains impossible at present due to the sign problem \cite{lQCD16,lQCD17}. Thermodynamics of QCD was also studied within the Functional Renormalization Group approach, which allowed to take under control quantum quark-meson fluctuations in the deep infrared limit \cite{FRG1,FRG2,FRG3,FRG4,FRG5,FRG6}. At the same time, a pure Yang-Mills potential for the Polyakov loop being an order parameter of the deconfinement transition does not provide a phase transition in a perturbative regime, see e.g. \cite{pLOOP1,pLOOP2,pLOOP3}. This difficulty can be overcome within an effective scalar theo\-ry whose minima correspond to the Polyakov loop expectation values \cite{effLOOP}. The discontinuous change of a global minimum provides the first order phase transition in the pure Yang-Mills case. The incorporation of this mechanism to the Nambu-Jona-Lasinio (NJL) model, which reproduces proper chiral dynamics \cite{NJL}, makes it possible to account for two of the most important aspects of QCD, e.g. the deconfinement of colour degrees of freedom and the dynamical restoration of chiral symmetry. 

In the past years this approach, known as the Polyakov-Nambu-Jona-Lasinio (PNJL) model, was successfully applied to the study of QCD thermodynamics\- at zero baryonic chemical potential \cite{PNJL1,PNJL2,PNJL3,PNJL4}. A systematic\- improvement of the model allowed it to account for effects of non-local current-current interaction \cite{PNJL5}, diquark degrees of freedom \cite{PNJL6} and meson-like correlations of quarks \cite{PNJL7}. At the same time, the  back-reaction of quarks propagating in a homogeneous temporal gluon field, which is assumed by the PNJL model, makes the Polyakov loop potential $\mathcal{U}$ dependent on $\mu_B$. The incorporation of this effect into the model has become an important step toward the understanding of the strongly interacting matter phase diagram, see e.g. \cite{PNJL8,PNJL9,PNJL10,PNJL11,PNJL12}. A perturbative estimate of this dependence \cite{PNJL8}, however, leads to identically zero $\mathcal{U}$ at $T=0$. As a result, the PNJL model at zero temperature\- does not encode information about the physical value of the Polyakov loop and, consequently, fails to reproduce its dynamics. Working out an improved parametrization\- of the Polyakov loop potential with a special emphasis on its density dependence is the primary goal of the present work.  Previous approaches to this problem have already been considered, see for example \cite{PNJL13}. The construction of a μB-dependent Polyakov loop potential leading to non artificial contribution of gluons to the baryonic density must be carefully considered.

The confined phase of the PNJL model should be identified with the hadronic one. Its description in the present model is rather schematic since, typically, it includes only scalar and pseudoscalar mesonic correlations of quarks \cite{PNJL7,PNJL8,PNJL11}. A more elaborate description of strongly interacting matter can be obtained within a hybrid EoS. Since the Polyakov loop is the deconfinement order parameter, it is natural to assume that its non zero expectation value suppresses hadronic degrees of freedom. Therefore, the EoS of hadron matter should switch to the PNJL EoS when the Polyakov loop attains a non zero value. In other words, the phase transitions given by the Gibbs criterion and defined by the order parameter should coincide. Below we consider a hybrid model with such a Polyakov-Gibbs phase transition.  We pay special attention to the case of electrically neutral $\beta$-equilibrated matter at zero temperature, which is of practical interest to astrophysical applications in neutron stars (NSs).  In our work we will deal with a standard treatment based on thermodynamical considerations although there are also some works where this may be externally triggered  \cite{exotic1,exotic2, exotic3}. 

The article is organized as follows. In the next section we briefly sketch the PNJL model. Section \ref{Potential} is devoted to the generalization of the Polyakov loop potential to the case of finite baryonic density. The hybrid quark-hadron EoS and the corresponding thermodynamic quantities of interest are discussed in section \ref{EoS}. Conclusions are given in section \ref{Concl}. 

\section{PNJL Model}
\label{PNJL}

In this work we consider the case of $N_f=3$ quark flavours with physical masses. We adopt the simplest form of the Lagrangian from Refs. \cite{PNJL2,PNJL3,PNJL4}, which provides dynamical restoration of chiral symmetry
\begin{eqnarray}
\label{I}
\mathcal{L}=\overline{q}(i\slashed{D}-\hat{m})q+
\frac{G}{2}\left[(\overline{q}q)^2+
(\overline{q}i\gamma^5\vec{\tau}q)^2\right]
-\mathcal{U}(\Phi,\Phi^*),
\end{eqnarray}
where the flavour space row $q=(\psi_u,\psi_d,\psi_s)^T$ stands for the quark field, the diagonal matrix $\hat{m}={\rm diag}(m_u,m_d,m_s)$ gives the corresponding masses and chiral-symmetric local four-point quark interaction in scalar and pseudoscalar channels is controlled by a coupling constant $G$.  The flavour mixing interaction channels of the 't Hooft determinant type \cite{tHooft} are neglected for the sake of simplicity. We, however, should note that, even if accounted, such terms do not significantly affect the thermodynamics of the present model but leading to a slight stiffening of its EoS \cite{HS2}. A covariant derivative $D^\mu=\partial^\mu-igA^\mu$ absorbs the static and uniform gluon field $A^\mu$, where $g$ is the gauge coupling. According to the PNJL model assumption, only the temporal component of this field has non zero value, i.e. in the case of three colours ($N_c=3$) one gets $A^\mu=\delta^\mu_0A^0_a\frac{\lambda^a}{2}$ with $\lambda^a$ being the Gell-Mann matrices. Within the present model the dynamics of gluons is reduced to the one of the Polykov loop. It is given in terms of temporal gauge fields as
\begin{eqnarray}
\label{II}
\Phi=\frac{1}{N_c}tr_c\left[
\mathcal{T}\exp\left(ig\int_0^\beta d\tau A^4\right)\right],
\end{eqnarray}  
where $\mathcal{T}$ is the time ordering operator, $\beta=\frac{1}{T}$ is the inverse temperature, $A^4=iA^0$. The involved trace in the previous expression is carried over colour indices. Hereafter all quantities are given in the natural system of units where the Boltzmann constant, the speed of light and the Planck constant are set $k_B=c=\hbar=1$. In the Polyakov gauge the temporal qluon field is diagonal in colour space and, thus, is controlled by only two independent non zero variables, $A^0_3$ and $A^0_8$ \cite{PNJL2}. This makes the Polyakov loop expectation value complex, i.e $\Phi^*\neq\Phi$ in the  general case. It, however, becomes real when quark chemical potentials vanish (see for example Ref. \cite{PNJL8} for a discussion).

The gluonic self-interaction of the non-abelian nature is modelled by the potential $\mathcal{U}(\Phi,\Phi^*)$. Its dependence on the Polyakov loop expectation value and corresponding complex conjugate is chosen as in Ref. \cite{PNJL4}. This choice provides center $Z(3)$ symmetry of $\mathcal{U}$ and its absolute mini\-mum at $|\Phi|=0$ or $|\Phi|\rightarrow1$ in the cases of small and high temperatures, respectively. Thus
\begin{eqnarray}
\label{III}
&&\mathcal{U}(\Phi,\Phi^*)=-\frac{b_2(T)}{2}\Phi^*\Phi 
\nonumber \\ 
&&\hspace*{.3cm}+b_4(T)\ln\bigl[1-6\Phi^*\Phi+4(\Phi^{*3}+\Phi^3)-3(\Phi^*\Phi)^2\bigl].\quad
\end{eqnarray}
Note that the logarithmic divergence of this potential appears as necessary to limit the expectation value of the Polyakov loop modulus from above. The medium dependent functions in Eq. (\ref{III}) are defined as 
\begin{eqnarray}
\label{IV}
b_2(T)&=& a_0T^4+ a_1T_0T^3+a_2T_0^2T^2, \\
\label{V}
b_4(T)&=& a_4T_0^3T.
\end{eqnarray}
A natural strategy to assign values of their parameters is to fit $\mathcal{U}$ to the lattice data of thermodynamics of pure gauge QCD since its pressure is $p_{glue}=- \mathcal{U}$. In Ref. \cite{PNJL4} this procedure gave $a_0=3.51$, $a_1=-2.47$, $a_2=15.22$, $a_4=-1.75$. Note, that in this case $T_0=270$ MeV represents the temperature of the deconfinement phase transition in the absence of quarks.

The present paper focuses on the zero temperature case, which is the most interesting for modelling evolved NSs with possible quark cores. The relevant  thermodynamic potential $\Omega$ can be obtained as a limit of the finite temperature case. Bosonization of Lagrangian in Eq. (\ref{I}) and a posterior mean field approximation is a standard procedure to obtain $\Omega$ as done in \cite{PNJL2,PNJL3,PNJL4,PNJL5}. An equivalent treatment is provided by the  introduction of the mean values of scalar $\langle\overline{q}q\rangle$ and pseudoscalar $\langle\overline{q}i\gamma^5\vec{\tau}q\rangle$ quark condensates along with a further linearisation procedure for $\mathcal{L}$ obtained from considering small deviations from these mean values. Thus
\begin{eqnarray}
\label{VI}
\frac{\Omega}{V}&=&\mathcal{U}(\Phi,\Phi^*)
+\frac{\langle\overline{q}q\rangle^2}{2G}
-\int_f\biggl[3(\omega_f^++\omega_f^-)\theta(\Lambda^2-{\bf p}^2)\nonumber\\
&+&2T\ln\left(1+3\Phi e^{-\beta\omega_f^+}
+3\Phi^* e^{-2\beta\omega_f^+}+
e^{-3\beta\omega_f^+}\right)\nonumber\\
&+&2T\ln\left(1+3\Phi^*e^{-\beta\omega_f^-}
+3\Phi e^{-2\beta\omega_f^-}+
e^{-3\beta\omega_f^-}\right)\biggl].\hspace*{.4cm}
\end{eqnarray}
Hereafter a symbolic notation for summation over all quark flavours and simultaneous integration over momentum $\int_f=\sum_f\int\frac{d^3{\vec p}}{(2\pi)^3}$ is introduced for shortening expressions. Single particle energies of quarks (superscript index ``$+$'') and antiquarks (superscript index ``$-$'')
\begin{eqnarray}
\label{VII}
\omega_f^\pm=\sqrt{{\bf p}^2+m_f^{*2}}\mp\mu_f,
\end{eqnarray} 
are defined through their effective masses
\begin{eqnarray}
\label{VIII}
m^*_f= m_f-G\langle\overline{q}q\rangle,
\end{eqnarray} 
and chemical potentials $\mu_f$. The latter ones are given in terms of quark baryonic charge $B_q=\frac{1}{3}$, electric charge of flavour $f$,  $Q_f$,  and associated baryonic $\mu_B$ and electric $\mu_Q$ chemical potentials
\begin{eqnarray}
\label{IX}
\mu_f=\mu_B B_q+\mu_Q Q_f.
\end{eqnarray}
Note, that the strange chemical potential $\mu_S$ is absent in Eq. (\ref{IX}) since the corresponding charge is not conserved if weak decays are allowed. The mean value of the scalar quark condensate $\langle\overline{q}q\rangle$ that minimizes the thermodynamic potential is defined by the condition
\begin{eqnarray}
\label{X}
\frac{\partial\Omega}{\partial\langle\overline{q}q\rangle}=0.
\end{eqnarray} 
Note, that the pseudoscalar quark condensate is absent in Eqs. (\ref{VI}) - (\ref{VIII}), since within the mean field approximation its mean value vanishes. 

A sharp momentum cut-off is introduced by using a  parameter $\Lambda$ to provide the simplest way of regularization of integrals as done in  \cite{PNJL2,PNJL3,PNJL4,PNJL6,PNJL12,PNJL13}. However, more refined regularization schemes implying smooth form factors have been also successfully applied in \cite{PNJL5,PNJL7,PNJL9}. In the present article this latter approach is not used for the sake of simplicity.  

Within the mean field approximation the Polyakov loop expectation value and its complex conjugate are defined by requiring the minimal value of the  thermodynamic potential, i.e. $\frac{\partial\Omega}{\partial\Phi}=\frac{\partial\Omega}{\partial\Phi^*}=0$. For known values of the Polyakov loop and scalar quark condensate, the pressure can be found as $p=-\frac{\Omega-\Omega_{vac}}{V}$. Here $\Omega_{vac}$ is the vacuum part of the thermodynamic potential. It does not contribute to the pressure and appears as the first term under the momentum integral in Eq. (\ref{VI}). Thus
\begin{eqnarray}
\label{XI}
p&=&2T\int_f\biggl[\ln\left(1+3\Phi e^{-\beta\omega_f^+}
+3\Phi^* e^{-2\beta\omega_f^+}+e^{-3\beta\omega_f^+}\right)
\nonumber\\
&+&\ln\left(1+3\Phi^*e^{-\beta\omega_f^-}
+3\Phi e^{-2\beta\omega_f^-}+
e^{-3\beta\omega_f^-}\right)\biggl]\nonumber\\
&-&\mathcal{U}(\Phi,\Phi^*)
-\frac{\langle\overline{q}q\rangle^2}{2G}.
\end{eqnarray}
It is worth noting that the vacuum term $\Omega_{vac}$ does not have any dependence on $\Phi$ and $\Phi^*$. From the previous  the expectation value of the Polyakov loop can be found from the conditions  
\begin{eqnarray}
\label{XII}
\frac{\partial p}{\partial\Phi}=
\frac{\partial p}{\partial\Phi^*}=0.
\end{eqnarray}
As it is seen from Eqs. (\ref{VI}) and (\ref{XI}), within the present model $\Phi$ and $\Phi^*$ are symmetrically coupled to quarks and antiquarks, respectively. Therefore, the asymmetry between them makes the Polyakov loop complex, while it becomes real only at $\mu_f=0$. Another source of the Polyakov loop dependence on quark chemical potentials as well as on flavour content is caused by the running of the QCD coupling. Within the PNJL model this dependence is accounted by the simple modification of the expansion coefficients $b_2$ and $b_4$ \cite{PNJL8,PNJL9,PNJL10,PNJL11,PNJL12}. More precisely, the back-reaction of quarks to the gluon sector leads to the modification of the transition temperature $T_0$ to become a function of $\mu_f$ and $N_f$. In Ref. \cite{PNJL8} a perturbative estimate of this dependence was done based on HDL/HTL results on the effective charge. 

At zero temperature, however, such an approach to account for the impact of quarks on the Polyakov loop is not satisfactory. First of all, the terms under the logarithms in Eq. (\ref{XI}), which explicitly couple $\Phi$ and $\Phi^*$ to quarks in expressions for thermodynamic potential and pressure, are exponentially suppressed at $T=0$. At the same time, the Polyakov loop potential in Eq. (3) with expansion coefficients $b_2$ and $b_4$ of the form shown in Eq.(\ref{IV}) and (\ref{V}) vanishes at zero temperature. This happens for any finite $T_0=T_0(\mu)$. Consequently, the modification of $\mathcal{U}$ motivated by the HDL/HTL perturbative estimate accounts for back-reaction of quarks to the gluon sector only in part. As a result, the dynamics of the Polyakov loop is totally lost in the PNJL model at zero temperature. Note, that gluons affect the properties of strongly interacting matter through quantum loops even at $T=0$, while their thermal excitations are suppressed in this case. Therefore, in the presence of dynamical quarks the Polyakov loop potential should explicitly depend on baryonic density even at zero temperature.

Before going any further we must discuss the normali\-zation of the present model. Typically, in the two-flavour case the coupling constant $G$ and cut-off parameter $\Lambda$ are fixed in order to reproduce vacuum values of the light quark condensate and effective quark mass, which is taken roughly equal to one third of the nucleon mass. Instead, for $N_f=3$ this normalization scheme requires a modification in order to account for effects of strange quarks. For this purpose we fitted the parameters of the present model to vacuum values of condensates of light $\langle\overline{l}l\rangle\equiv\langle\overline{u}u+\overline{d}d\rangle/2$ and strange $\langle\overline{s}s\rangle$ quarks. Our used vacuum value of $\langle\overline{l}l\rangle^{1/3}$ coincides within the error bars with the recent result $\langle\overline{l}l\rangle^{1/3}= 283(2)$ MeV from lattice simulations in three-flavour QCD \cite{QC}. While $\langle\overline{s}s\rangle^{1/3}$ exceeds the lattice value 290(15) MeV by about 9 \%, that when accounting for the errors bars reduces this deviation to 3 \%. As we discuss latter, this fine tuning of the input data was done in order to take under control the speed of sound in the deconfinement region. The pion decay constant $f_\pi$ is obtained from the well-known Gell-Mann-Oakes-Renner relation $\frac{f_\pi^2m_\pi^2}{2}=-\frac{m_u+m_d}{2}\langle\overline{l}l\rangle_0$ with the discussed condensate of light quarks and physical masses of quarks and pion \cite{PDG}. It is worth noting, that, remarkably, this $f_\pi$ value is very close to the most recent one reported by the Particle Data Group $f_\pi=130.2(1.7)$ MeV \cite{PDG}. The set of model parameters, vacuum condensates, current quark masses used, as well as mass and decay constant of pion are listed in Table \ref{table1}.

The present set up gives $m^*_u=182.6$ MeV, $m^*_d=185.1$ MeV and $m^*_s=275.4$ MeV in the vacuum. These values are smaller than $\sim$ 300 MeV and $\sim$ 500 MeV usually accepted for light and strange quarks, respectively. Such a difference is caused by the rather schematic quark interaction in the present model, which accounts only for scalar and pseudoscalar channels. These masses can be taken under control by introducing the 't Hooft determinant interaction channel, which is omitted in order to keep the quark sector of the model as simple as possible. It is also appropriate to note here, that we do not use the parameter set of Ref. \cite{PNJL4} since it was found for the two flavour case, while for three flavours it gives too large constituent masses in vacuum being $\sim$ 700 MeV and $\sim$ 800 MeV for light and strange quarks, respectively.

\begin{table}[!]
\begin{tabular}{|c|c|c|c|c|c|c|c|c|c|c|c|}
\hline
 $m_u$ [MeV] & $m_d$ [MeV] & $m_s$ [MeV] & $~\Lambda$ [MeV] & $G$ [GeV$^{-2}$]\\ \hline                       
     2.2     &     4.7     &    95.0     &      925.06     &     2.385         \\ \hline    
\end{tabular}
\begin{tabular}{|c|c|c|c|c|c|c|c|c|c|c|c|}
\hline
 $|\langle\overline{l}l\rangle_0|^{1/3}$ [MeV] & 
 $|\langle\overline{s}s\rangle_0|^{1/3}$ [MeV] & $f_\pi$ [MeV] & $m_\pi$ [MeV]\\ \hline 
          281           &       315            &     126.96    &    139.3     \\ \hline
\end{tabular}
\caption{Parameters of the present model (top row) and resulting physical quantities (bottom row).}
\label{table1}
\end{table}
%

\section{Density dependent Polyakov loop potential}
\label{Potential}

As it was mentioned in the previous section, terms coupling $\Phi$ and $\Phi^*$ to quarks in Eqs. (\ref{VI}) and (\ref{XI}) are exponentially suppressed at small temperatures. Therefore, within the original PNJL model, the equations for the Polyakov loop become $\frac{\partial\mathcal{U}}{\partial\Phi}=\frac{\partial\mathcal{U}}{\partial\Phi^*}=0$. For the parameters of the Polyakov loop potential from Ref. \cite{PNJL4} these equations have non zero solution only at temperatures exceeding 263 MeV. At small temperatures $\Phi$ is identically equal to zero, which formally should be interpreted as a confinement of colour charge at all baryonic densities. This, however, contradicts the existing phenomenology of QCD. Even a HDL/HTL motivated perturbative modification of $T_0$, which becomes a function of chemical potential, does not resolve this paradox. It can be solved, however, by introducing an additional dependence of the Polyakov loop potential $\mathcal{U}$ on the quark chemical potential or, alternatively, on the baryonic density. The technical advantage of the latter will become evident in the following. At the same time, from the physical point of view, such a treatment is totally equivalent to the case, when $\mathcal{U}$ depends on $\mu_f$ as done in Ref. \cite{PNJL13}. The simplest way to perform the discussed generalization of the Polyakov loop potential is to leave its dependence on $\Phi$ and $\Phi^*$ the same as in Eq. (\ref{III}), while making functions $b_2$ and $b_4$ dependent on baryonic density. For this purpose we propose a simple parametrization 
\begin{eqnarray}
\label{XIII}
b_2(T,n_B)\hspace*{-.2cm}&=&\hspace*{-.2cm} 
a_0T^4+ a_1T_0T^3+a_2T_0^2T^2+
\widetilde{a}_2T_0^4\left(\frac{n_B}{T_0^3}\right)^{\kappa_2}
\hspace*{-.3cm},\hspace*{.5cm} \\
\label{XIV}
b_4(T,n_B)\hspace*{-.2cm}&=&\hspace*{-.2cm} 
a_4T_0^3T+
\widetilde{a}_4T_0^4\left(\frac{n_B}{T_0^3}\right)^{\kappa_4},
\end{eqnarray}
with $\widetilde{a}_l$ and $\kappa_l\ge0$ ($l=2,4$) being constants defined below. With such a modification of $\mathcal{U}$ the Polyakov loop can have a finite value even at zero temperature.

Within the present model $\Phi$ is considered as an effective scalar field representing gluonic degrees of freedom and treated under the mean field approximation. Typically, such a mean field framework leads to the shifting of the single particle energies giving rise to an effective chemical potential
\begin{eqnarray}
\label{XV}
\mu_f^*\equiv\mu_f+B_q\mathcal{V}(\Phi,\Phi^*).
\end{eqnarray}
Here $\mathcal{V}$ is some unknown function, which is supposed to be the same for all quark flavours under the assumption that their interaction with gluons is universal. We also require $\mathcal{V}=0$ at $n_B=0$ to ensure baryon-antibaryon symmetry in this limit.

Up to here we considered the finite temperature case. Let us now focus on the zero temperature case, keeping in mind that all main results are general and applicable at $T\neq0$. Hereafter we also explicitly introduce electrons to the consideration since they are important to neutralize quark matter, which is the most interesting for astrophysical applications. Their pressure $p_e$ is nothing else as the one of noninteracting spin-$\frac{1}{2}$ fermions with mass $m_e=0.511$ MeV \cite{PDG} and chemical potential $\mu_e=-\mu_Q$. With the above modifications of the Polyakov loop potential and effective chemical potentials of quarks $\mu_f^*$, the zero temperature pressure expression in the PNJL model reads
\begin{eqnarray}
\label{XVI}
p=-6\int_f\omega_f^{*+}\theta(-\omega_f^{*+})
-\mathcal{U}(\Phi,\Phi^*)
-\frac{G}{2}\langle\overline{q}q\rangle^2+p_e.
\end{eqnarray}
Here $\omega_f^{*+}$ is the modified single particle energy of the form given in Eq. (\ref{VII}), where the physical chemical potential $\mu_f$ is replaced\- by the effective one $\mu_f^*$, while the quark condensate and Polyakov loop are still defined by Eqs. (\ref{X}) and (\ref{XII}), respectively.

At this point we focus on $\mathcal{V}$, which parametrizes the present model. However, its choice  is not arbitrary. This follows from the analysis of the baryonic density defined as $n_B=-\frac{1}{V}\frac{d\Omega}{d\mu_B}$. The conditions $\frac{\partial\Omega}{\partial\Phi}=\frac{\partial\Omega}{\partial\Phi^*}=0$ and $\frac{\partial\Omega}{\partial\langle\overline{q}q\rangle}=0$ significantly simplify its calculation giving
\begin{eqnarray}
n_B=-\frac{1}{V}\left[\frac{\partial\Omega}{\partial\mu_B}+
\frac{\partial\Omega}{\partial n_B}
\frac{d n_B}{d \mu_B}\right]=
\frac{\partial p}{\partial\mu_B}+
\frac{\partial p}{\partial n_B}\frac{d n_B}{d \mu_B},\,
\end{eqnarray}
where the relation between the thermodynamic potential and pressure was used on the second step. Finally, with the help of Eqs. (\ref{XV}) and (\ref{XVI}) the baryonic density expression  becomes
\begin{eqnarray}
\label{XVII}
n_B&=&6B_q\int_f\theta(-\omega_f^{*+})\nonumber\\
&+&\left[6B_q\int_f\theta(-\omega_f^{*+})
\frac{\partial\mathcal{V}}{\partial n_B}
-\frac{\partial\mathcal{U}}{\partial n_B}\right]
\frac{\partial n_B}{\partial\mu_B}.~
\end{eqnarray} 
The first term in this expression corresponds to the contribution of quarks, while the second one includes derivatives of potentials $\mathcal{U}$ and $\mathcal{V}$ associated with the Polyakov loop. This means that the second term can be connected to gluons that, as known, do not carry baryonic charge and can not contribute to $n_B$. This paradox becomes even more evident in absence of dynamical quarks i.e when their masses approach the infinitely heavy limit. In this case the momentum integrals in Eq. (\ref{XVII}) vanish, while baryonic density still retains a  finite value $n_B=-\frac{\partial\mathcal{U}}{\partial\mu_B}$. In other words, the introduction of some dependence of the Polyakov loop potential on baryonic chemical potential or baryonic density would spuriously lead to baryonic charge of gluons. In order to solve this problem we require that the square bracket expression  in Eq.  (\ref{XVII}) is zero so that in this case baryonic density equals to
\begin{eqnarray}
\label{XVIII}
n_B=6B_q\int_f\theta(-\omega_f^{*+}).
\end{eqnarray} 
Consequently, this leads  to the condition
\begin{eqnarray}
\label{XIX}
n_B\frac{\partial\mathcal{V}}{\partial n_B}-
\frac{\partial\mathcal{U}}{\partial n_B}=0,
\end{eqnarray} 
which relates the potentials $\mathcal{U}$ and $\mathcal{V}$. It is important to stress that at finite temperature the condition (\ref{XIX}) written in terms of the baryonic density remains exactly the same, while expression for $n_B$ itself obviously gets modified by thermal excitations of quarks. The fulfillment of this relation ensures that at any baryonic chemical potential and temperature only quarks contribute to baryonic density. The above analysis also shows why it is more convenient from the practical point of view to consider the discussed potentials as functions of baryonic density instead of baryonic chemical potential.  We can easily check that Eq. (\ref{XIX})  is fulfilled by the function
\begin{eqnarray}
\label{XX}
&&\hspace*{-.4cm}\mathcal{V}(\Phi,\Phi^*)=
-\frac{c_2(n_B)}{2}\Phi^*\Phi \nonumber\\ 
&&\hspace*{.2cm}+c_4(n_B)\ln\bigl[1-6\Phi^*\Phi+4(\Phi^{*3}+\Phi^3)
-3(\Phi^*\Phi)^2\bigl],\hspace*{.6cm}
\end{eqnarray}
which preserves the same dependence on the Polyakov loop as in the case of the potential $\mathcal{U}$ and involves coefficients
\begin{eqnarray}
\label{XXI}
c_l(n_B)=\frac{\kappa_l b_l(0,n_B)}{(\kappa_l-1)n_B}.
\end{eqnarray} 
For further convenience we also introduce the notation
\begin{eqnarray}
\label{XXII}
d_l(n_B)=n_Bc_l(n_B)-b_l(0,n_B)=\frac{b_l(0,n_B)}{\kappa_l-1}.
\end{eqnarray}

As it follows from Eqs. (\ref{XX}) and (\ref{XXI}), the potential $\mathcal{V}$ vanishes at $n_B=0$ for an arbitrary temperature only if either $\kappa_l=0$ or $\kappa_l>1$. An additional limitation on the possible values of $\kappa_l$ can be found from the analysis of the regime of high baryonic densities, when effective masses of quarks $m_f^*$ become negligible compared to their chemical potentials $\mu_f^*$. In this case fermionic contribution to the pressure, i.e. the first term in Eq. (\ref{XVI}), behaves as $n_B^{4/3}$, while the second one, which comes from the Polyakov loop potential, is proportional to $n_B^{\kappa_2}$ for $\kappa_2\ge\kappa_4$ and $n_B^{\kappa_4}$ otherwise. The term associated with the quark condensate can be neglected. Since at high baryonic densities quarks are expected to be massless and asymptotically free \cite{afree}, the total pressure should behave as $p \sim \mu_B^4$. It follows from the thermodynamic identity $n_B=\frac{\partial p}{\partial\mu_B}$ that such an asymptotic $p$ is obtained only if $\kappa_l\le\frac{4}{3}$. At the same time, the requirement of the baryonic charge-anticharge symmetry can be fulfilled only if $b_l$ is an even function of baryonic density. Along with the other limitations on possible values of $\kappa_l$ this result yields either $\kappa_l=0$ or $\frac{4}{3}$. As it is shown below, the mean value of the Polyakov loop modulus approaches one at high densities only if $\frac{b_4}{b_2}\rightarrow0$. Therefore, we set $\kappa_2=\frac{4}{3}$ and $\kappa_4=0$. In the asymptotic case of high densities this corresponds to $b_2\sim\mu_B^4$ and constant $b_4$.  Note that the contribution of gluons to baryonic density coming from the derivative $\frac{\partial\mathcal{U}}{\partial\mu_B}$ is compensated.

At this point, the Polyakov loop potentials $\mathcal{U}$ and $\mathcal{V}$ have two free parameters, i.e. $\widetilde{a}_2$ and $\widetilde{a}_4$. The value of $\widetilde{a}_2$ can be estimated by assuming that the ratio of the symmetric quark matter pressure to the Stefan-Boltzmann pressure $p_{SB}$ is the same in the limits of infinite temperature and baryonic chemical potential, i.e. $\frac{p}{p_{SB}}\bigl|_{\mu_B\rightarrow\infty}=\frac{p}{p_{SB}}\bigl|_{T\rightarrow\infty}=\alpha_{SB}$. The present assumption is supported by the consistency of lattice data on the QCD EoS for 2, 2+1 and 3 quark flavours \cite{lQCD1} and $\mathcal{O}(\alpha_s^2)$ perturbative calculations at zero temperature \cite{pQCD}. In both of these cases $\alpha_{SB}=0.8$ within estimated error bars.  We also should note that at present, unfortunately, there is no reliable information about QCD matter at high densities. Thus, relaying on the perturbative results seem to be the only available approach to normalize the model.

At this regime the modulus of the Polyakov loop approaches unity and the logarithmic term in the expression for $\mathcal{U}$ can be neglected. Therefore, $\mathcal{U}\simeq-\frac{a_2}{2}n_B^{4/3}$ in this case. The quark contribution to the pressure is the one of $N_f$ species of massless non-interacting fermions with spin-colour degeneracy $6$, i.e. $\frac{1}{4}\bigl(\frac{\pi^2}{N_f}\bigl)^{1/3}\bigl(\frac{n_B}{B_q}\bigl)^{4/3}$. The finite contribution being quadratic in the quark condensate can be neglected. Then, with the help of the thermodynamic identity $n_B=\frac{\partial p}{\partial\mu_B}$ the dependence of the total pressure on baryonic density can be turned to the one on baryonic chemical potential. This yields 
\begin{eqnarray}
\label{XXIII}
\frac{p}{p_{SB}}\Bigl|_{\mu_B\rightarrow\infty}=
\alpha_{SB}=\left(1+2\widetilde{a}_2B_q^{\frac{4}{3}}
\left(\frac{N_f}{\pi^2}\right)^{\frac{1}{3}}\right)^{-3},
\end{eqnarray}
where the Stefan-Boltzmann pressure $p_{SB}=\frac{N_f(B_q\mu_B)^4}{4\pi^2}$ was used. Finally, using this expression we obtain
\begin{eqnarray}
\label{XXIV}
\widetilde{a}_2=\frac{1}{2B_q^{\frac{4}{3}}}\left(\frac{\pi^2}{N_f}\right)^{\frac{1}{3}}\left(\alpha_{SB}^{-\frac{1}{3}}-1\right)=
0.358\cdot N_f^{-\frac{1}{3}}.
\end{eqnarray}
For $N_f=3$ this gives $\widetilde{a}_2\simeq0.25$.

\begin{figure}[t]
\includegraphics[width=0.9\columnwidth]{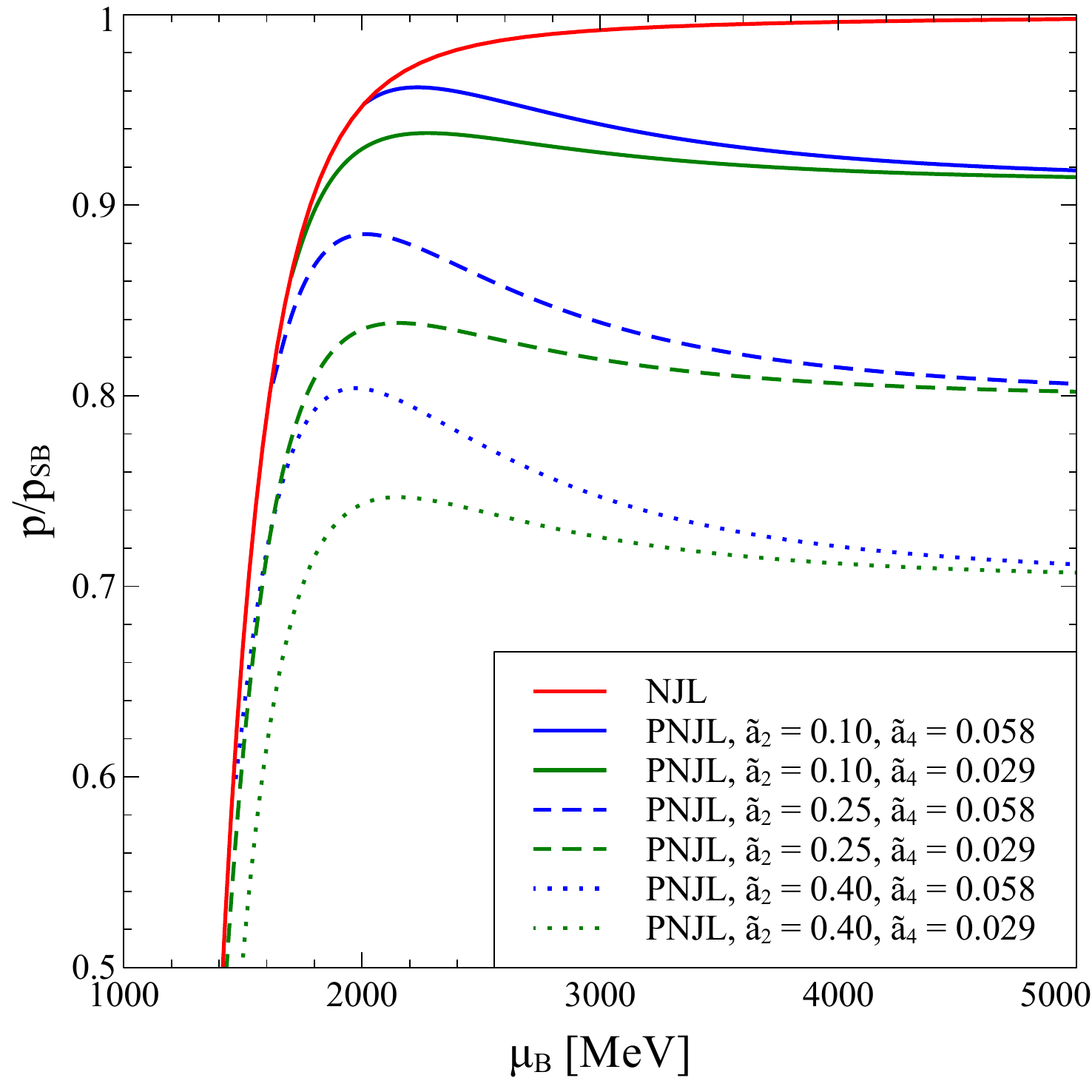}
\caption{Scaled pressure of three flavour symmetric quark matter as a function of baryonic chemical potential $\mu_B$ at different values of the Polyakov loop potential parameters.}
\label{fig1}
\end{figure}

In order to numerically check this result we calculated the zero temperature pressure of three flavour symmetric ($\mu_Q=0$) quark matter as function of baryonic chemical potential $\mu_B$ for different values of coefficients $\widetilde{a}_2$ and $\widetilde{a}_4$. As it is seen from Fig. \ref{fig1}, the value of $\widetilde{a}_2$, indeed, provides correct asymptotic of the pressure corresponding to about 80\% of the Stefan-Boltzmann pressure. This result holds for any value of $\widetilde{a}_4$. At the same time, NJL model without Polyakov loop potential significantly overestimates the pressure compared to results of perturbative calculations.

 The present model is an effective low energy approximation of QCD and, strictly speaking, is not applicable to the analysis of thermodynamics at infinite density. We assume that its applicability range is limited by $\mu_B$ at which quark chemical potential becomes comparable to the cut off parameter, i.e. by $\mu_B\simeq3\mu_f=3\Lambda\simeq3$ GeV. From the practical point of view we are interested only in chemical potentials $\mu_B\lesssim2$ GeV reached inside the NSs \cite{HS2}, which is well inside the estimated applicability range of the present model.

At zero temperature the calculation of the Polyakov loop expectation value is significantly simplified since in this case $\Phi$ and $\Phi^*$ enter the expression for the pressure Eq. (\ref{XVI}) only through potentials $\mathcal{U}$ and $\mathcal{V}$. Then, by direct calculation it is possible to show that $\Phi\frac{\partial p}{\partial\Phi}-\Phi^*\frac{\partial p}{\partial\Phi^*}\sim\Phi^3-\Phi^{*3}$. On the other hand, this expression equals to zero due to requirement (\ref{XII}). This means that the expectation value of Polyakov loop belongs to the SU(3) center subgroup, i.e. $\Phi=|\Phi|e^{i\frac{2\Pi l}{3}}$ with $l=0,1,2$. As a result, in the zero temperature case potentials $\mathcal{U}$ and $\mathcal{V}$ depend only on the modulus of the Pokyakov loop expectation value, which can be found from Eq. (\ref{XII}) as
\begin{eqnarray}   
\label{XXV}
|\Phi|=\frac{1}{3}+\frac{2}{3}\sqrt{1+\frac{9d_4(n_B)}{d_2(n_B)}}.
\end{eqnarray}
Besides this solution the Eq. (\ref{XII}) also posseses the trivial root $|\Phi|=0$, which corresponds to the confinement state. 

\section{Equation of state}
\label{EoS}

Expression (\ref{XXV}) demonstrates that $|\Phi|$ approaches unity at high densities only if $\frac{d_4}{d_2}=-\frac{b_4}{3b_2}\rightarrow0$ in full agreement with values found for $\kappa_2$ and $\kappa_4$. On the other hand, this limiting value of the Polyakov loop modulus should be reached from below. This means, that $\frac{d_4}{d_2}<0$. According to the definition (\ref{XXII}) and value of $\widetilde{a}_2$ this is the case only for $\widetilde{a}_4>0$. Therefore, the solution in Eq. (\ref{XXV}) is meaningful only if the expression under the square root is positive. This allows us to conclude that within the present model the modulus of the Polyakov loop is non zero only at baryonic densities larger than a critical value 
\begin{eqnarray}   
\label{XXVI}
n_c=
\left(\frac{3\widetilde{a}_4}{\widetilde{a}_2}\right)^{\frac{3}{4}} T^3_0,
\end{eqnarray}
while $|\Phi|=0$ for $n_B<n_c$. It is also seen from Eq. (\ref{XXV}) that $|\Phi|=\frac{1}{3}$ at $n_B=n_c$. Treating the Polyakoov loop expectation value as the deconfinement order parameter we should assume that quark matter exists only at $n_B\ge n_c$ when $|\Phi|$ has non zero values. At the same time, at smaller densities $|\Phi|=0$ and strongly interacting matter exists in the form of hadrons. This gives us a phenomenological criterion to construct within the present model a hybrid quark-hadron EoS.

\begin{figure}[t]
\includegraphics[width=0.9\columnwidth]{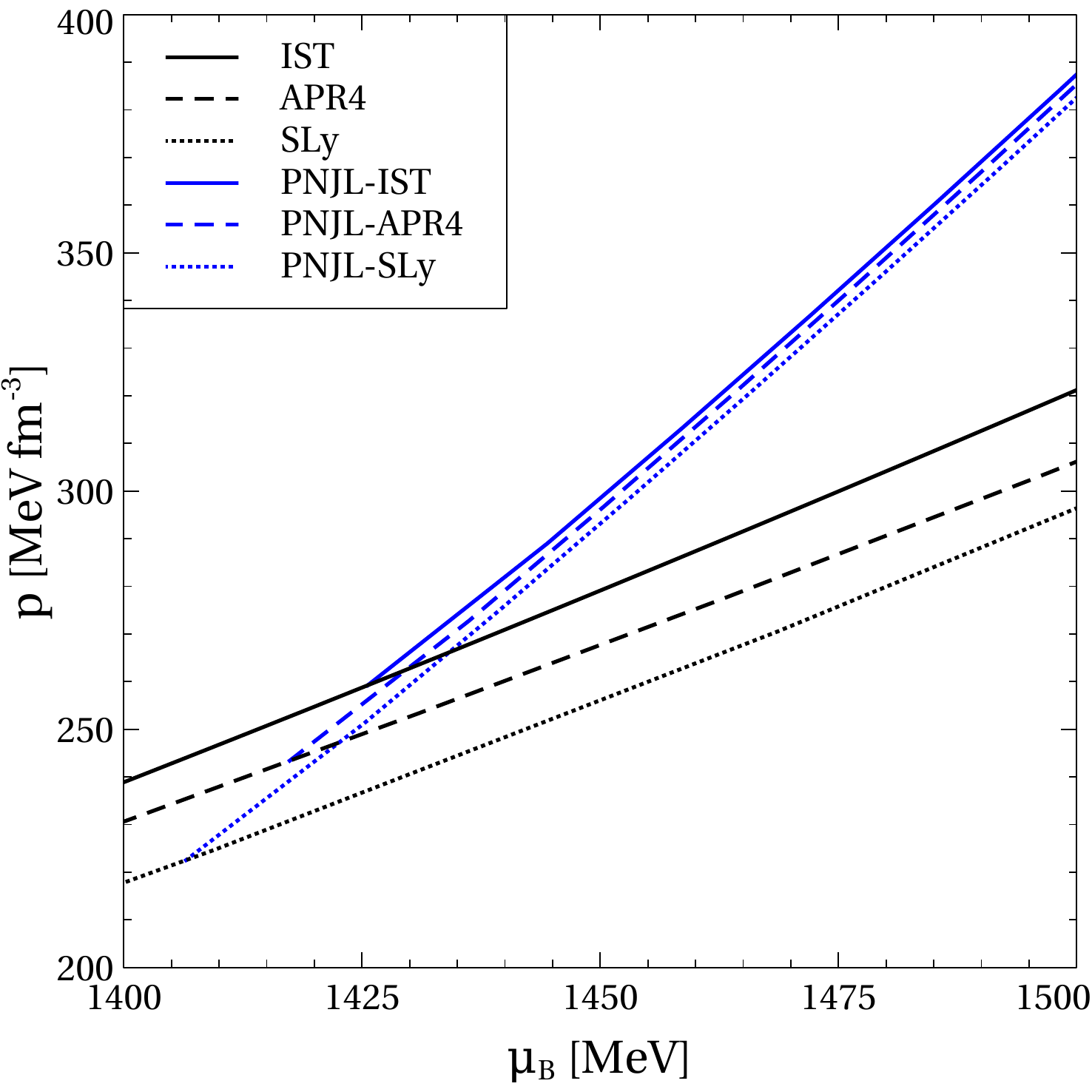}\\
\includegraphics[width=0.9\columnwidth]{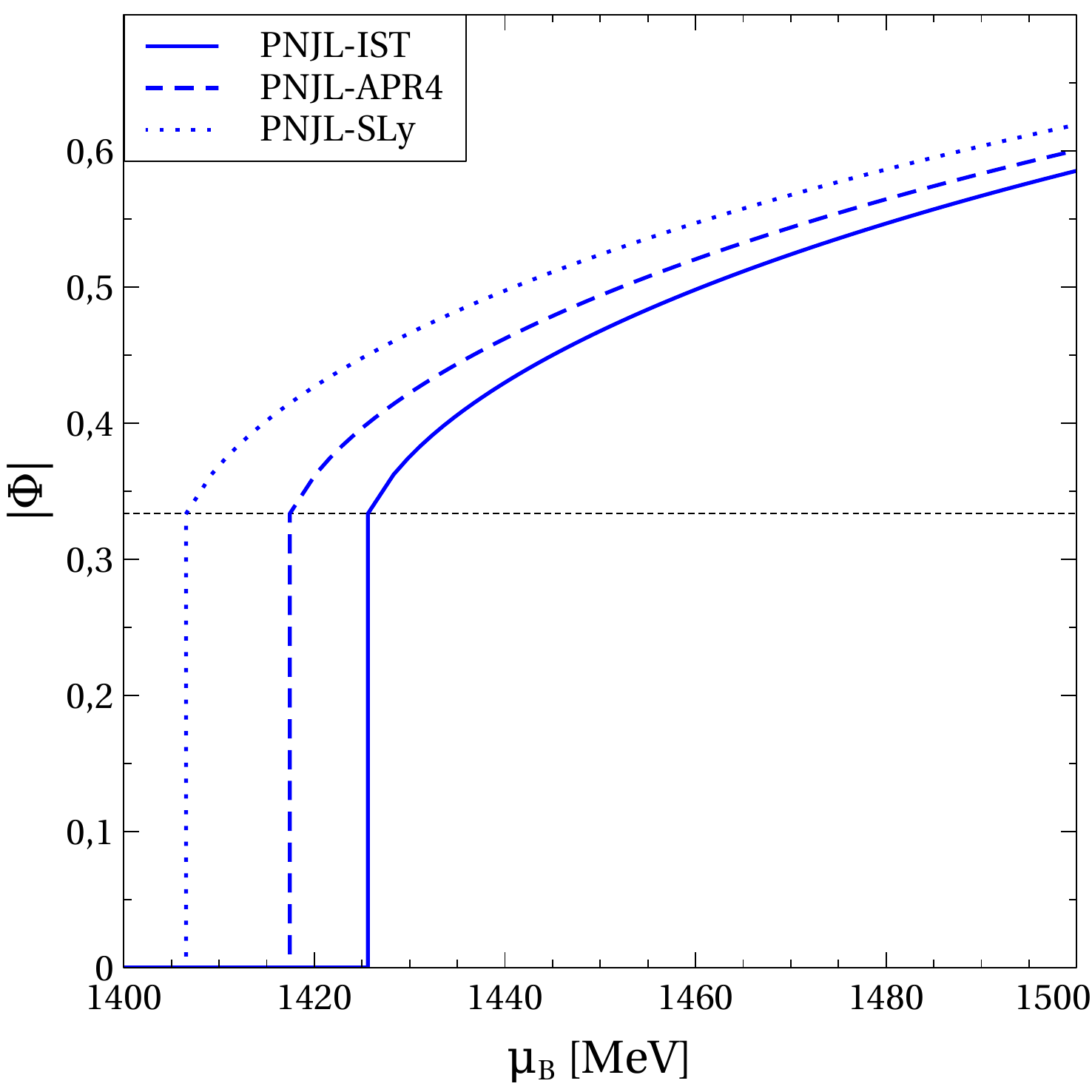}
\caption{Pressure $p$  (upper panel) and Polyakov loop modulus $|\Phi|$ (lower panel) of electrically neutral matter as a function of baryonic chemical potential $\mu_B$ in the phase transition region. Black dotted line on the lower panel represents $|\Phi|=\frac{1}{3}$.}
\label{fig2}
\end{figure}

For this purpose we utilize the Gibbs construction which requires that the pressures and chemical potentials of the two phases coincide at the phase transition. This corresponds to the dynamical and chemical equilibrium of phases. Thermal equilibrium is trivially provided since we consider the case of zero temperature. We also consider electrically neutral matter since it is the most interesting for astrophysical applications. Electric chemical potential $\mu_Q$ is defined by the condition of zero total density of electric charge, i.e. from
\begin{eqnarray}
\label{XXVII}
n_Q=6\int_f\theta(-\omega_f^{*+})Q_f-n_e=0,
\end{eqnarray} 
where $n_e=\frac{\partial p_e}{\partial \mu_e}$ defines the particle density of electrons. Note that the equilibrium with respect to $\beta$-decay is provided automatically since $\mu_d=\mu_u+\mu_e$ by construction.

The coefficient $\widetilde{a}_4$ plays an important role in the construction of the resulting hybrid EoS. It is chosen to satisfy the condition that the modulus of the Polyakov loop receives non zero value $|\Phi|=\frac{1}{3}$ exactly at the quark phase boundary as obtained by the Gibbs criterion. In other words, the definition of phase transition  given by the Gibbs criterion and the behaviour of the order parameter coincide in the present model. It is clear that the parameters of quark-hadron phase transition and the corresponding value of $\widetilde{a}_4$ depend on the particular hadronic EoS. 

In this work we use three hadronic EoSs. Two of them, i.e. the APR4 EoS \cite{APR4}, which stands for the parameterization of the microscopic potential A18+$\delta$v+UIX, and the SLy EoS \cite{SLy}, are usually used as references in many nuclear and astrophysical studies. The third hadronic EoS, the IST \cite{IST} is able to fulfil many experimental and observational constraints on properties of nuclear and hadron matter. It is necessary to note that the mentioned hadronic EoSs do not include strangeness content as it only appears in quark matter due to weak processes converting $d$ and $s$ quarks to each other. Matching hadronic EoSs with the developed procedure in the present paper we obtained three hybrid EoSs labelled below as PNJL-IST, PNJL-APR4 and PNJL-SLy, respectively.

\begin{table}[t]
\begin{tabular}{|l|c|c|c|c|c|c|c|c|c|c|c|}
\hline
         & $\widetilde{a}_4$ &$n_B^h$ [fm$^{3}$]&$n_B^q$ [fm$^{3}$] \\ \hline 
PNJL-IST &     0.032     &        0.80      &      1.25         \\ \hline 
PNJL-APR4&     0.031     &        0.73      &      1.23         \\ \hline 
PNJL-SLy &     0.030     &        0.75      &      1.19         \\ \hline 
\end{tabular}
\caption{Values of $\widetilde{a}_4$, baryonic densities of pure hadronic $n_B^h$ and quark $n_B^q$ phases coexisting at the deconfinement phase transition.}
\label{table2}
\end{table}
\begin{figure}[t]
\includegraphics[width=0.9\columnwidth]{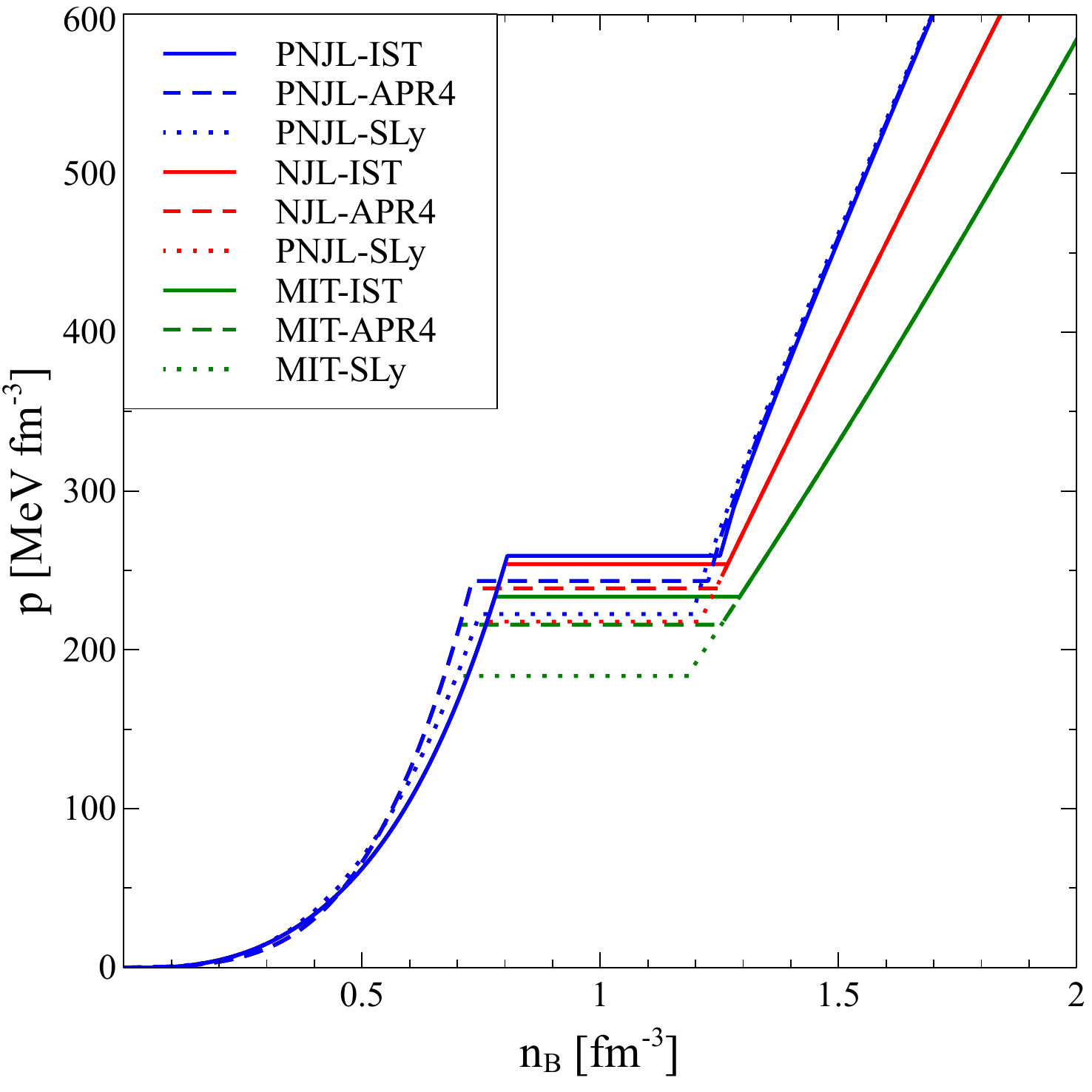}\\
\includegraphics[width=0.9\columnwidth]{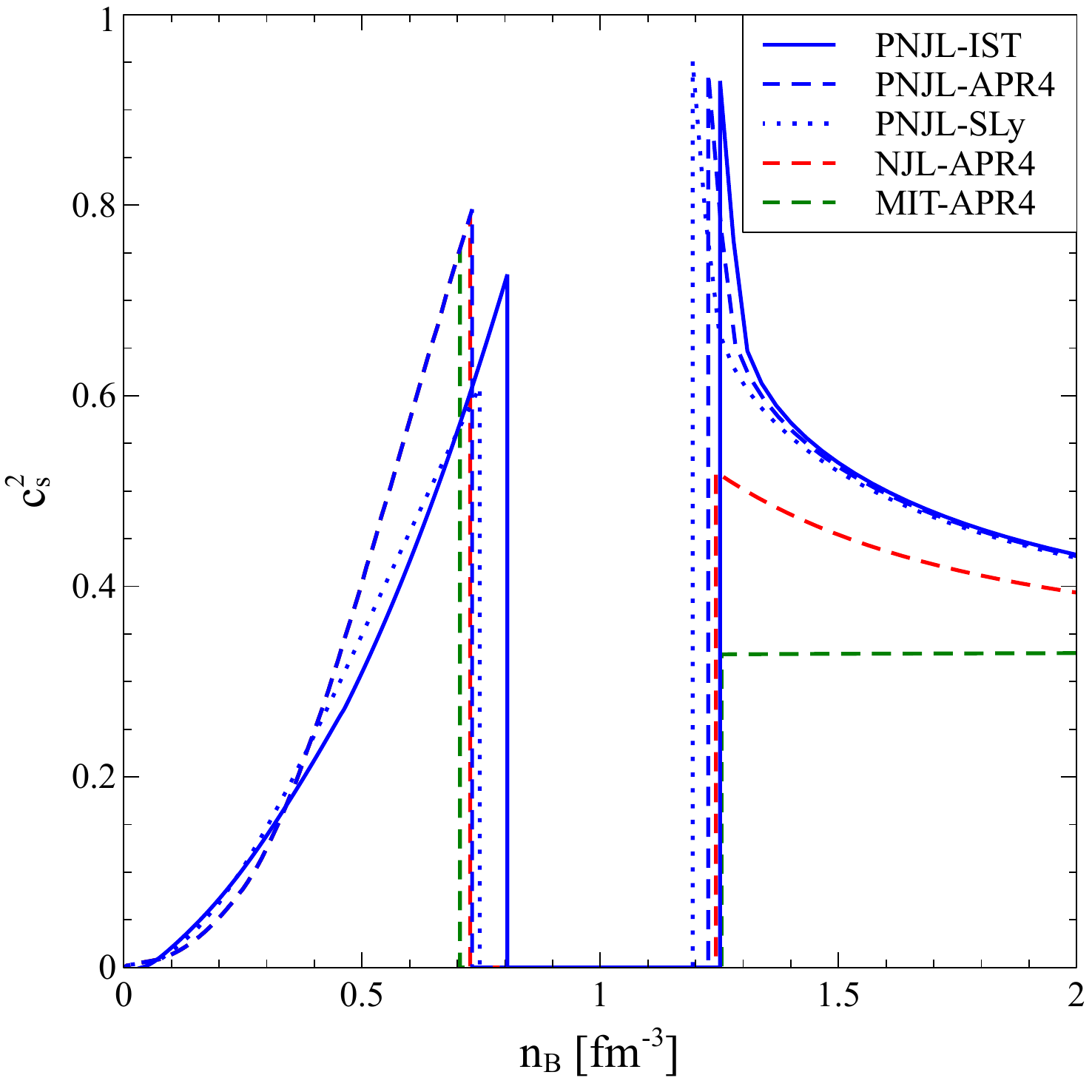}
\caption{Pressure $p$ (upper panel) and squared speed of sound $c_{\rm s}^2$ (lower panel) of electrically neutral matter as a function of baryonic density  $n_B$ calculated for hybrid EoS with the quark phase described by the PNJL (blue curves), NJL (red curves) and MIT bag (green curves) models.}
\label{fig3}
\end{figure}

The upper panel of Fig. \ref{fig2} shows the pressure of electrically neutral quark phase for each of these hybrid EoSs (blue curves) as a function of baryonic chemical potential in the  the phase transition region. The transition from hadronic matter (black curves) happens  when the pressures of two phases coincide. As is seen from the lower panel of Fig. \ref{fig2}, which shows the Polyakov loop modulus versus baryonic chemical potential, before this transition $|\Phi|=0$ indicating a hadron phase of the strongly interacting matter. Quark matter exists above the transition point, where the modulus of the Polyakov loop attains non zero value approaching unity at high $\mu_B$. A discontinuous jump of the Polyakov loop modulus at the phase transition reveals that $|\Phi|$, indeed, is the deconfinement order parameter. The corresponding values of parameter $\widetilde{a}_4$ together with  baryonic densities of coexistence at the deconfinement transition hadron $n_B^h$ and quark $n_B^q$ phases are given in Table \ref{table2}. As it is seen from Fig. \ref{fig2} the behaviour of the quark matter pressure is weakly sensitive to details of hadronic EoS used to find $\widetilde{a}_4$. This is reflected by the close values found for baryonic densities of $n_B^h$ and $n_B^q$ corresponding to the PNJL-IST, PNJL-APR4 and PNJL-SLy EoSs. The same conclusion can be drawn from the upper panel of Fig. \ref{fig3}, which depicts the pressure of electrically neutral matter at zero temperature as a function of baryonic density. Indeed, the quark sector of the hybrid EoSs constructed with the PNJL model and different hadronic EoS (blue curves) are barely distinguishable by eye. It worth noting, that it happens the same case for hybrid EoSs constructed with the NJL model without the Polyakov loop (red curves) and the MIT bag model (green curves) with the bag constant $B^{1/4}=200$ MeV, which are shown for comparison. The quark EoS of the PNJL model is sizably stiffer than the one of the NJL model. This feature of the present model can provide positive feedback to the two solar mass limit problem for NSs \cite{2sol} or even for a  disconnected third-family branch of compact stars in the mass-radius relationship \cite{tobias}.

Within the NJL-inspired models this problem can also be resolved by considering a phenomenological vector interaction producing universal repulsion between quarks \cite{vecNJL1}. Such a vector interaction stiffens an EoS and corresponds to a quadratic behaviour of pressure at high baryonic densities, i.e. $p\sim n_B^2$ at $n_B\rightarrow\infty$. This leads to $p\sim\mu_B^2$ being inconsistent with results of perturbative QCD $p\sim\mu_B^4$ \cite{pQCD,pQCDold}. We, however, may think on NJL-inspired models as a low energy approximation which should not necessarily reproduce the high density behaviour of QCD. At the same time, the baryonic chemical potential in the compact star interiors can reach values up to 2 GeV \cite{HS2}, which is well inside the estimated applicability range of the present model. Zero temperature  $\mathcal{O}(\alpha_s)$, $\mathcal{O}(\alpha_s^2)$ and partial $\mathcal{O}(\alpha_s^3\ln^2\alpha_s)$ perturbative results show that at such $\mu_B$ scaled pressure $\frac{p}{\mu_B^4}$ differs from the corresponding asymptotic value by just a few percent \cite{pQCD,pQCDold}. This means that at the highest densities typical for the compact star interiors the behaviour of pressure is close to $p\sim\mu_B^4$. Therefore, the ability of the present model to provide such a behaviour of EoS is important for astrophysical applications. In addition, this EoS predicts the phase transition onset at $n_B$ roughly equal to 4-5 normal nuclear densities $n_0$, which is significantly larger than $\sim2.5n_0$ in models with vector quark interaction \cite{HS2}. Note that such a small density of deconfinement is also reported in other works \cite{vBAG,paramQEOS}. They, however, either explicitly include repulsive vector interaction \cite{vBAG} or use parametrisation of the deconfined phase EoS, which leads to $p\sim n_B^{c_s^2+1}$ with $c_s^2=0.8$ or $1$ \cite{paramQEOS} being very close to the quadratic form produced by vector interaction channels.

We also paid a special attention to the behaviour of the speed of sound of the present model defined as $c^2_{\rm s}=\frac{dp}{d\epsilon}$ with $\epsilon$ being an energy density. The dependence of this quantity on $n_B$ is shown on the lower panel of Fig. \ref{fig3}. First, we note that at high densities $c_{\rm s}^2$ approaches a limiting value $\frac{1}{3}$ which is provided by the pressure asymptote  $p\sim\mu_B^4$. At the same time, the speed of sound is a decreasing function of baryonic density. In the case of the NJL model without Polyakov potential $c_{\rm s}^2$ also decreases with $n_B$, while for the MIT bag model it is constant due to the absence of the quark masses and interaction. In the right vicinity of phase transition the speed of sound of the present model has quite a  large value $\sim0.9$. Moreover, at some particular values of coupling $G$ and cut off parameter $\Lambda$ it can even become superluminal. In this work $c_{\rm s}^2<1$ was provided by the adjustment of the above mentioned parameters. This explains why our vacuum value of the strange quark condensate is slightly larger than the lattice one. In fact, the vacuum value of $\langle\overline{s}s\rangle^{1/3}$ is just 3 \% larger than the lattice value on the upper edge of its error bars \cite{QC}. Taking into account that the quark interaction in the present model is rather schematic, we find the reached consistency with the lattice data more than satisfactory. We expect that the introduction of additional realistic interaction channels will make this consistency even better and it is left for future work. At the same time, it is necessary to stress that a superluminal speed of sound appears even in very advanced effective theories. See e.g. Fig. 25 of Ref. \cite{HS2}, where this problem was resolved by adjusting the coupling constants which control the quark interaction in vector and s-wave, spin-singlet, flavor- and color-antitriplet channels. Moreover, the asymptotic value of $c_{\rm s}^2$ predicted by that work exceeds $\frac{1}{3}$ due to the presence of vector interaction channel leading to $p\sim\mu_B^2$ and $c_{\rm s}^2\rightarrow1$ at high densities. 

\section{Conclusions}
\label{Concl}

We re-examined the PNJL model at finite baryonic densities in order to incorporate the Polyakov loop dynamics at zero temperature being of interest to  modelling NSs with quark cores. The main question addressed in this work is how the Polyakov loop potential depends on baryonic density or, equivalently, baryonic chemical potential. We demonstrated that typically used HDL/HTL perturbative estimate of this dependence, unfortunately, is inapplicable at zero temperature since it leads to zero value of the deconfinement order parameter at all baryonic densities. In order to solve this problem we performed a  phenomenological generalization of the Polyakov loop potential to the case of finite $n_B$. The introduction of an arbitrary dependence of $\mathcal{U}$ on $n_B$ or $\mu_B$ can be  formally interpreted as originating the presence of baryonic charge of gluons. This paradox appears at all temperatures and is the most evident in the absence of dynamical quarks when their current masses approach infinity. As we show it can be solved by introducing the Polyakov loop dependent shift of a single quark energy, which is absorbed by an effective chemical potential of quarks, in the spirit of the mean field framework of the Polyakov loop treatment within the PNJL model. We derived a relation between the  corresponding single particle potential $\mathcal{V}$ and the Polyakov loop potential $\mathcal{U}$, whose fulfillment ensures an absence of gluonic contribution to the baryon charge density. 
Moreover, it has the same form regardless of the quark interaction channels present in the model nor any particular form of the Polyakov loop potential $\mathcal{U}$. We expect that the cancellation of gluonic contribution to the baryon charge density at finite temperatures can be important for a reliable \-modelling of the QCD phase diagram. 

The analysis of the present model asymptotic behaviour at high baryonic densities provided us with a very tight restriction on a possible dependence of $\mathcal{U}$ on $n_B$. In fact, the  uncertainty remaining corresponds only to unknown values of two constant parameters, i.e. $\widetilde{a}_2$ and $\widetilde{a}_4$. Furthermore, based on our model EoS and results of the $\mathcal{O}(\alpha^2_s)$ perturbative calculations at zero temperature we found that in the case of  three quark flavours $\widetilde{a}_2=0.25$. The remaining free parameter of the density dependent Polyakov loop potential $\widetilde{a}_4$ was used in order to match EoSs of quark and hadron matter at the deconfinement phase transition. We used the Gibbs criterion together with the requirement that  the Polyakov loop jumps exactly at the phase transition. Such an approach to construct a phase transition in hybrid quark-hadron EoS simultaneously provides the existence of chemical and dynamical equilibrium of coexisting phases as well as a discontinuous behaviour of its order parameter. We used three different EoSs of hadron matter and all them yielded roughly the same density of the deconfinement onset around $4-5~n_0$. At the same time we should note that the onset of the deconfinement can be shifted to smaller densities if a quark-hadron phase boundary is not sharp but smoothed due to small values of surface tension. 

As an important consequence of non zero values of the Polyakov loop at zero temperature the stiffening of the model EoS arises. Technically this effect is caused by a contribution coming from the Polyakov loop potential $\mathcal{U}$. As it appears it can provide the NS interiors with an ability to resist the gravitational collapse and, consequently, to reach the well-known two solar mass limit for NSs. At the same time, such a stiffening\- leads to the increase of the speed of sound. In fact, it has the maximal value in the right vicinity of the deconfinement phase transition. A further increase of the baryonic density leads to the  decrease of the speed of sound, which has asymptotic value $c_s^2=\frac{1}{3}$ of free massless quarks in a full agreement with asymptotic freedom of quarks expected at high baryonic densities. We also expect that non zero values of the Polyakov loop can affect properties of colour superconducting phase, which should be studied in the future.

\section*{Acknowledgments}

We are grateful to K. Bugaev, D. Blaschke and T. Fischer for valuable comments. The  work of OI, MAPG and CA  was  performed within the project SA083P17 funded by Junta de Castilla y Le\'on (Spain). They also acknowledge the financial support from University of Salamanca. VS acknowledges the financial support through the grant UID/FIS/04564/2019 of the Funda\c c\^ ao para a Ci\^ encia e  Tecnologia (FCT) of Portugal.


\end{document}